\documentclass[twocolumn,floatfix,amsmath,prb,aps,showpacs]{revtex4}

\usepackage{amsmath}
\usepackage{amssymb}
\usepackage{graphicx}
\usepackage{epsfig}
\usepackage{psfrag}

\begin{document}

\title{Wigner crystal {\em vs.} Friedel oscillations in the 1D Hubbard model}

\author{Stefan A.~S\"{o}ffing}
\author{Michael Bortz}
\author{Imke Schneider}
\author{Alexander Struck}
\author{Michael Fleischhauer}
\author{Sebastian Eggert}
\affiliation{Dept.~of Physics and Research Center OPTIMAS, Univ.~Kaiserslautern, D-67663 Kaiserslautern, Germany }

\date{\today}

\newcommand{\ddk}{\Delta_k}
\newcommand{\dk}{\tfrac{\pi}{L+1}}
\newcommand{\kf}{k_{\text F}}
\newcommand{\todo}[1][]{(#1 ???) \marginpar{$\nLeftarrow$}}
\newcommand{\be}{\begin{eqnarray}}
\newcommand{\ee}{\end{eqnarray}}
\def\refeq#1{(\ref{#1})}
\def\d{\mbox d}
\def\wt{\widetilde}
\def\nn{\nonumber}
\def\i{\int_{-\infty}^{\infty}}
\def\ip{\int_{0}^{\infty}}
\def\mi{\int_{-\infty}^{0}}
\def\up{\uparrow}
\def\down{\downarrow}
\def\l{\left}
\def\r{\right}
\def\rmi{{\rm{i}}}
\def\d{\mbox d}
\def\i{\int_{-\infty}^{\infty}}
\def\ip{\int_{0}^{\infty}}
\def\mi{\int_{-\infty}^{0}}
\def\te{\mbox{e}}

\begin{abstract}

We analyze the fermion density of the one-dimensional Hubbard model  
using  bosonization and numerical DMRG calculations.  
For finite systems we find a relatively sharp crossover 
even for moderate short range interactions into a
region with $4\kf$ density waves as a function of density.
The results show that the unstable fixed point of a spin-incoherent state can 
dominate the physical behavior in a large region of parameter space in finite 
systems.
The crossover may be observable in ultra cold fermionic gases in optical lattices
and in finite quantum wires.
\end{abstract}


\maketitle

\section{Introduction}\label{intro}

Having been predicted in the early days of quantum mechanics, the
 Wigner crystal\cite{wigner} is one of the simplest but most dramatic many-body effects:
Due to the long range repulsive forces,
electrons spontaneously form a self-organized lattice 
at low enough densities and temperatures 
much different from a free electron gas.
Experimental verification has been difficult, but
very recently signatures of a Wigner crystal were
reported in carbon nanotubes.\cite{bockrath}
Using ultra cold gases in optical lattices it is now also possible
to produce well-controlled correlated fermion systems in restricted 
dimensions,
albeit with {\it  short-range} interactions.\cite{esslinger,blochrev}

The Wigner crystal in one dimension (1D) has been discussed so far mostly in the 
context of long-range interactions.  In that case it has been 
predicted by Schulz \cite{schulz1} that the density-density 
correlations corresponding to an
equally spaced inter-particle distance become dominant.
Numerically a crossover to Wigner density waves has been observed at strong
long-range interactions with very few particles.\cite{dots}
However, for {\it short-range} interacting fermion systems in 
1D possible Wigner crystal signatures 
have not been addressed theoretically so far.

The prototypical model for short-range interacting fermions is the
repulsive ($U>0$) 1D Hubbard model
\begin{equation}
	H = -t \sum_{ x=1}^{L-1} \left( \psi_{\sigma, x}^\dagger 
\psi_{\sigma, x+1}^{\phantom{\dagger}} + h.c. \right) 
		+ U \sum_{x=1}^L n_{\uparrow, x} n_{\downarrow, x}.
	\label{ham}
\end{equation}
Even though many exact results have been derived for this model
using the Bethe Ansatz,\cite{book} the local densities in finite 
chains with open boundaries cannot yet be calculated by exact methods.

It is known that at low energies 
the short range interacting system in Eq.~(\ref{ham}) becomes effectively scale invariant 
up to well understood logarithmic corrections in the spin-channel. 
Therefore, 
any crossover towards a different physical region would be unexpected.
Moreover, a Wigner crystal region should be unstable, since 
the $2\kf$-density Friedel oscillations \cite{friedel} are always the
slowest decaying correlations due to a bounded Luttinger parameter\cite{schulz1}
$0.5\leq K_c\leq 1$. 
Nonetheless, sub-dominant oscillations at other wave-numbers also 
exist as has been explicitly shown 
e.g.~for the Hubbard model in a finite magnetic field.\cite{bedurftig,sirker}

We now study the density distribution in finite Hubbard chains with hard-wall
boundaries by a combination of bosonization and numerical 
density matrix renormalization group (DMRG) calculations. 
In Sec.~\ref{dens} we describe how $2k_F$ and $4 k_F$ density 
oscillations typically arise in 1D systems.
The corresponding analytical expressions  for the model (\ref{ham})
are derived using bosonization in Sec.~\ref{bos}.  Using the 
bosonization results it is then possible to accurately analyze the numerical results in 
Sec.~\ref{num}.  The physical interpretations and conclusions are 
presented in Sec.~\ref{concl}.
Despite the fact that the
interactions are short ranged and of moderate strength,
we find that a region with dominant $4\kf$  oscillations is 
always stable at low filling.
The results show that the scale invariance is explicitly broken.
The observed Wigner crystal region illustrates that
in 1D even short range interactions
have an increasing effect with {\it growing} inter-particle spacing.
The observed crossover is related to  the so-called
spin-incoherent Luttinger liquid.\cite{matveev}

\section{Density Oscillations} \label{dens}

The density profile in a system of interacting particles can be 
taken as a good indication of the nature of the ground state.  For 
example, let us consider a true Wigner crystal in 1D with broken translational 
symmetry, where the particles are localized at regular distances.
If the average filling is given by $n_0$,
the corresponding density $n(x)$ would then be well described by a
sum of Gaussian wavefunctions localized at positions $x=j/n_0$
with some localization length $\xi$
\be
n(x) & = & \sum_{j=-\infty}^{\infty} \frac{1}{\sqrt{2 \pi}\xi}
\exp\left(-\frac{(x- j /n_0)^2}{2 \xi^2} \right) 
\nonumber \\
& = & n_0 \, \, \theta_3\!\left(\pi n_0 {x}, e^{-2 \pi^2 \xi^2n_0^2} 
\right) \label{density1}
\ee
where we have used an exact identity for the elliptic Jacobi Theta function 
$\theta_3(x,q) \equiv \sum_j q^{j^2}e^{2 j i x}$ on the second line.  

Interestingly, for moderate particle distance in relation to the
localization length $1/n_0 \alt 3 \xi$, we have
$e^{-2 \pi^2 \xi^2n_0^2} \ll 1$ and the  expression (\ref{density1})
 is well approximated by a simple oscillation of the form
\be
n(x) \approx n_0 \left(1 + 2 e^{-2 \pi^2 \xi^2n_0^2} \cos(2 \pi n_0 x) \right).
\label{wigner}
\ee
Since the Fermi point of a filled Fermi sea with density $n_0$ 
is given by $\kf = n_0 \frac{\pi}{2}$ in the thermodynamic limit, 
it is clear that the oscillations in Eq.~(\ref{wigner}) correspond to a 
$4\kf$ density wave.  In this sense, $4\kf$ density 
oscillations and the Wigner crystal 
state can be taken to be equivalent phenomena.
The amplitude of the oscillations in Eq.~(\ref{wigner}) can even be used to extract the
effective localization length $\xi$.

\begin{figure}
   \begin{center}
        \includegraphics[height=0.50\textwidth,angle=270]{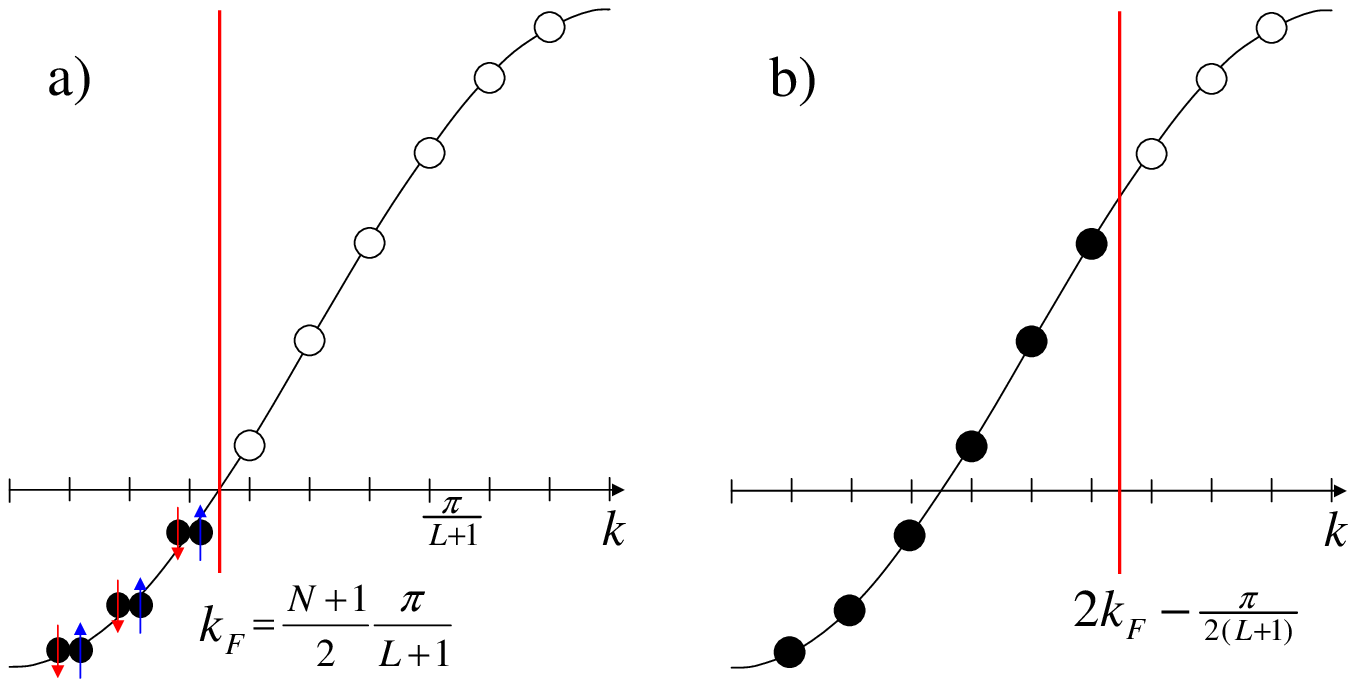}
   \end{center}
    \caption{Schematic occupied states 
for a) the nearly free case and b) for strong 
interactions ("large Fermi sea").}
\label{filling}
\end{figure}

There is, however, another possible reason for $4\kf$ oscillations in a finite system,
which is linked to open boundary conditions and
the arrangement of standing waves in momentum space.  Due to 
interactions fermions of opposite spin may 
no longer be allowed to occupy the same single particle states, and the Fermi sea is filled
to effectively twice the original Fermi point
relative to the noninteracting case as shown in Fig.~\ref{filling}.
The summation over standing waves $\sin( n  \dk  x)$
results in a density of the form 
\begin{eqnarray}
    n(x) 
       &=& \frac{2}{L+1} \sum_{n=1}^{N} \sin^2(n \dk x) 
\nonumber\\  &  =&  n_0
-\frac{\sin\left( 4 \kf  -\dk  \right)x}{2(L+1)  \sin \left(\dk x \right) } ,
\label{wigner2}
\end{eqnarray}
where we have used ''open'' boundary conditions $\psi_{\sigma,0}=\psi_{\sigma,L+1}=0$
on a finite lattice with a lattice constant of unity.
The Fermi point $\kf=\frac{N+1}{2}\dk $ is centered between
the highest occupied and lowest unoccupied level in a system with $N$ non-interacting 
fermions 
with spin as indicated in Fig.~\ref{filling}.

The density of the large Fermi sea in Eq.~(\ref{wigner2}) again shows 
$4\kf$ density oscillations, but in contrast to the true Wigner crystal 
in Eq.~(\ref{wigner}) the translational invariance is broken only by the open boundary 
conditions.  There is no spontaneous symmetry breaking in the large Fermi sea, 
since the amplitude in Eq.~(\ref{wigner2}) vanishes in the thermodynamic limit. 
Nonetheless, the $4\kf$ oscillations indicate a localization of particles near the boundary
with a localization length
that can be estimated by comparing the amplitude with  $2 n_0 e^{-2 \pi^2 \xi^2n_0^2}$
from Eq.~(\ref{wigner}). Indeed, 
the Wigner crystal state and the large
Fermi sea are strongly related in 1D.  Both can only arise due to strong 
repulsive interactions and both are  {\it spin incoherent} states, which are characterized
by a (nearly) degenerate spin sector.   The Wigner crystal state has been used as a 
starting point to derive the concept of a spin-incoherent Luttinger liquid  in the limit of
very strong long range interactions by allowing 
additional fluctuations.\cite{matveev}  On the other hand the large Fermi sea is in fact 
the {\it exact} spin-degenerate ground state of the Hubbard
model in Eq.~(\ref{ham}) in the limit of $U\to \infty$ as can be shown by the Bethe Ansatz.
{\it Therefore, both states represent fixed points of infinitely 
 strong interactions, which are unstable for any finite interaction strength.}
In this sense the amplitude of the $4\kf$ oscillations plays the role of an order
parameter which should be vanishing in the thermodynamic limit
in either case.  However, in finite systems or at finite energy scales it is in 
principle possible to observe regions that are dominated by unstable fixed points.
In fact as we will see from the numerical simulations,  
the order parameter remains finite in systems with open boundary conditions defining a 
region with dominant 
spin-incoherent behavior, which is separated by a well-defined crossover. 
~\\

In the non-interacting limit it is well-known that a finite system 
exhibits $2\kf$ Friedel oscillations,\cite{friedel} which are calculated by summing over 
double occupied  standing waves as shown in Fig.~\ref{filling}-a
\begin{eqnarray}
    n(x) 
       &=& \frac{4}{L+1} \sum_{n=1}^{N/2} \sin^2(n \dk x) 
\nonumber\\  &  =&  
n_0 - 
\frac{\sin\left( 2 \kf  x \right)}{(L+1)  \sin \left(\dk x \right) } ,
\label{friedel}
\end{eqnarray}
It has been predicted that the Friedel oscillations decay slower due to 
interactions,\cite{egger} but as our numerical results show the 
amplitude is  in fact strongly
suppressed with increasing $U$.  It must be emphasized 
here that
the $4\kf$ oscillations discussed above
are not simply a higher harmonic of the Friedel oscillations, but 
are an independent interaction effect which in fact competes with the
Friedel oscillations as we will see later.

\section{Bosonization results} \label{bos}
Using standard bosonization\cite{egg07} it is possible
to predict how finite interactions modify the analytic form of the 
$2\kf$ and $4\kf$ oscillations.
In particular, in the low energy effective theory after 
linearization around $\kf$ of the fermion fields $\psi_{\sigma,x} 
\approx e^{i\kf x}\psi_{R,\sigma} + e^{-i\kf x}\psi_{L,\sigma}$
one can identify 
the Friedel oscillations as an expectation value of the 
bosonic operator\cite{boundaries,egger,rommer}
\be
{\cal O}_{LR}  & = &\l(  e^{i2 \kf x}\psi^\dagger_{L,\sigma}(x)
\psi^{\phantom{\dagger}}_{R,\sigma}(x)+h.c. \r)
\nonumber \\ & \propto & 
 \sin (2 \kf x + \sqrt{2\pi K_c} \varphi_c)\cos(\sqrt{2\pi} \varphi_s).
\ee
Here the spin and charge fields represent the mode expansions $\nu=s,c$
\be
\varphi_\nu(x)&=& \tfrac{Q}{\sqrt{K_\nu}} \tfrac{x}{L+1}
+\sum_{n=1}^L \tfrac{1}{\sqrt{\pi n}} \sin\l(\tfrac{\pi n}{L+1}x\r)
\l( a_{n,\nu} +a_{n,\nu}^\dagger\r)
\label{mode}
\ee
according to open boundary conditions $\psi_L(x)=-\psi_R(-x)$.\cite{boundaries}
The number counting operator $Q$ does not contribute in ground state
expectation values $\langle Q\rangle =0$.
Using the standard calculations of correlation functions in finite systems
\cite{modecalc} the expectation value
is determined to be
\be \langle {\cal O}_{LR}\rangle  \propto  
\frac{\sin\left( 2 \kf x \right)}{\left[(L+1)  \sin
\left(\dk x \right)\right]^{(K_c+1)/2} } , \label{2kF}
\ee
up to logarithmic corrections.\cite{schulz0}
The interactions change the
decay rate of the Friedel oscillations compared to Eq.~(\ref{friedel}),
which appear to be {\it enhanced} for repulsive interactions $K_c <1$.
However, as we will see, the yet undetermined amplitude is strongly suppressed
with interactions and at low fillings.

The derivation of the 
Wigner oscillations from bosonization is more subtle.
They arise from interactions because of the
Umklapp term in the Hamiltonian density 
\be
{\cal O}_U & = & g_3 \left(e^{i 4 \kf x} \psi_{R,\uparrow}^\dagger 
\psi_{L,\uparrow}^{\phantom\dagger}
\psi_{R,\downarrow}^\dagger \psi_{L,\downarrow}^{\phantom\dagger} +h.c.\right)
\nonumber \\
&   \propto &     \cos \left(4 \kf x +\sqrt{8 \pi K_c} \varphi_c\right), \label{OU}
\ee
where $g_3\propto U$.  In first order perturbation theory this operator induces a
density expectation value 
$\langle n\rangle_U= \langle n\rangle-\langle n\rangle_0$
relative to the unperturbed case 
\be
\langle n(x)\rangle_U  \propto  \int_0^L dy \sum_\alpha
\frac{{}_0\langle  0|{\cal O}_U(y)|\alpha\rangle\langle\alpha| \partial 
\varphi_c(x)|0 \rangle_0 }{E_\alpha-E_0}
\ee
where $|\alpha\rangle$ are all excited states.  
By using the mode expansion in Eq.~(\ref{mode})
it is possible to 
calculate the expectation values for all bosonic excitations\cite{modecalc} 
$|\alpha\rangle$
with the result 
\be
\langle n(x)\rangle_U & \propto & 
  \frac{g_3 K_c}{v_c} 
\int_0^L \frac{\sin{(4  \kf y)} g(x,y)}{\left[(L+1)\sin
\left(\dk y\right)\right]^{2K_c}}\d y,
\ee
where 
 $g(x,y)=\sum_{m=1}^L\tfrac{1}{m} \sin\l(\frac{\pi m }{L+1}y\r) 
\cos\l(\frac{\pi m }{L+1}x\r) \approx \tfrac{\pi}{2}\theta(y-x)$.
Using $\int_x^\infty \sin(4\kf y) y^{-2K_c} \d y  \approx 
\cos{4 \kf x} /4\kf x^{2K_c} +{\cal O}(x^{-2K_c-1})$
the integral can be approximated  as\cite{footnote}
\be
\langle n(x)\rangle_U & \propto & 
 \frac{g_3 K_c}{v_c \kf} \frac{\sin(4 \kf -\dk) x}{\left[(L+1)\sin
\left(\dk x\right)\right]^{2K_c}}. \label{4kF}
\ee
The decay rate for the $4\kf$ oscillations is faster than for the Friedel oscillations
in Eq.~(\ref{2kF}) since $K_c \ge 0.5$ for the Hubbard model.
The linear dependence on $g_3/\kf \propto UL/N$ is only accurate to lowest order
in perturbation theory, but the 
typical oscillatory behavior and powerlaw will describe the 
behavior for any $U$ and filling $N/L$.
Alternatively, the {\it ad-hoc} inclusion of ${\cal O}_U$ directly
in the operator expression for the density is also a valid approach.\cite{schulz1}
The explicit derivation from perturbation theory above
now provides additional information
by indicating an increase of the amplitude
with $g_3/\kf$, i.e.~with larger interactions and smaller filling.
Note, that both the Friedel oscillations in Eq.~(\ref{2kF}) and the
$4\kf$ oscillations in Eq.~(\ref{4kF}) have a  non-zero expectation value only 
in systems with open boundary conditions.
However, the exact amplitude cannot be derived from bosonization, so that 
numerical calculations have to be used. 

\begin{figure}
   \begin{center}
        \includegraphics[width=0.50\textwidth,angle=0]{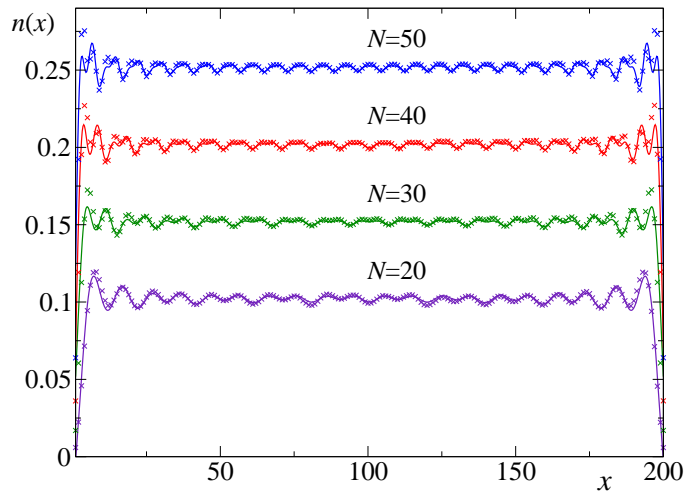}
   \end{center}
    \caption{Local density for $U=4t$ and $L=200$
for different filling 
showing the Friedel and Wigner crystal oscillations.  The solid lines correspond
to the theoretical prediction in Eq.~(\ref{rho_fit}).}
\label{density}
\end{figure}

\begin{figure}
   \begin{center}
        \includegraphics[width=0.50\textwidth,angle=0]{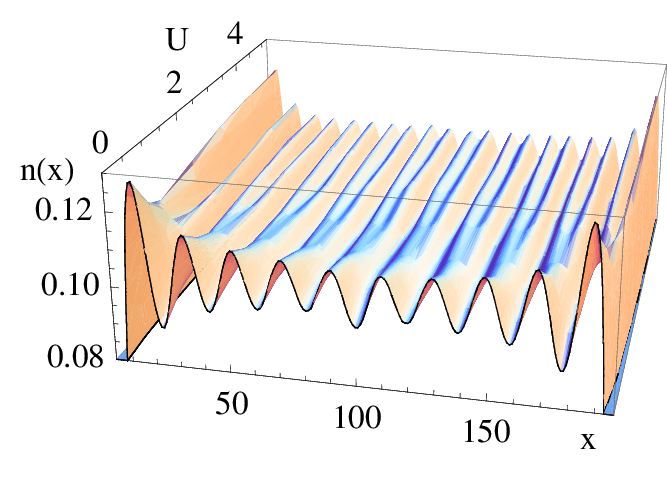}
   \end{center}
    \caption{Local density for $N=20$ and $L=200$ showing the
  crossover from $2\kf$ to $4\kf$ oscillations   with increasing interaction
strength $U$.  }
\label{density2}
\end{figure}

\section{Numerical Results} \label{num}

We have implemented a DMRG algorithm \cite{White1992} for the model in Eq.~(\ref{ham}) in order to
calculate the local density in finite systems with a given fermion number $N$.
 Typical densities at various fillings  are shown in
Fig.~\ref{density} for $U=4t$ and $L=200$, 
which clearly exhibit the predicted oscillations.  Figure \ref{density2} shows
how the local density at a given filling of $N/L=0.1$ emerges from the slower Friedel
oscillations to a Wigner crystal pattern as $U$ increases.

\begin{figure}
   \begin{center}
        \includegraphics[width=0.50\textwidth,angle=0]{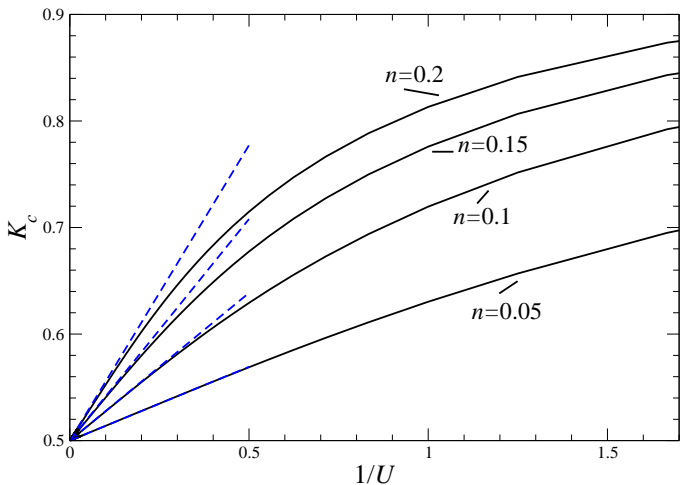}
   \end{center}
    \caption{The Luttinger liquid parameter $K_c$ for the Hubbard model
from Bethe Ansatz compared to the asymptotic form in Eq.~(\ref{kc}) for
different fillings.}
\label{ba}
\end{figure}

An accurate data analysis is now possible in terms of our 
analytic predictions from Eqs.~(\ref{2kF}) and (\ref{4kF})
\begin{eqnarray}
n(x) 
	\!	= \! n_0 
	\!	- \!A_1 \frac{\sin\left( 2\kf x \right)} {
\left[ 
\sin\left(\dk x \right) \right]^{\frac{K_c+1}{2}}}
	\!	- \!A_2 \frac{\sin\left( 4\kf\! -\!\dk  \right)x }{\left[ 
\sin \left(\dk x \right) \right]^{2 K_c}}
		\label{rho_fit}
	,
\end{eqnarray} 
where the Friedel amplitude $A_1$ and the Wigner amplitude $A_2$ can be
determined from fits to the numerical data.
For arbitrary interactions $U>0$ and filling $N/L$
the Luttinger parameter $0.5 \le K_c \le 1$ can be calculated 
exactly\cite{book,schulz1,sirker}
as shown in Fig.~\ref{ba},
so that the amplitudes
in the middle of the chain $A_1$ and $A_2$ are in fact the 
only two adjustable fitting parameters.
For the non-interacting case in Eq.~(\ref{friedel}) we have
 $A_1=\frac{1}{L+1}$ and $A_2=0$,
which we can use as a test of the numerical accuracy.
The uniform density $n_0$ is fixed by the requirement that $\int n(x) dx =N$,
so that it is not an independent fitting parameter.
(e.g. $n_0= \frac{N}{L+1}$ for $U=0$).

\begin{figure}
   \begin{center}
        \includegraphics[width=0.50\textwidth,angle=0]{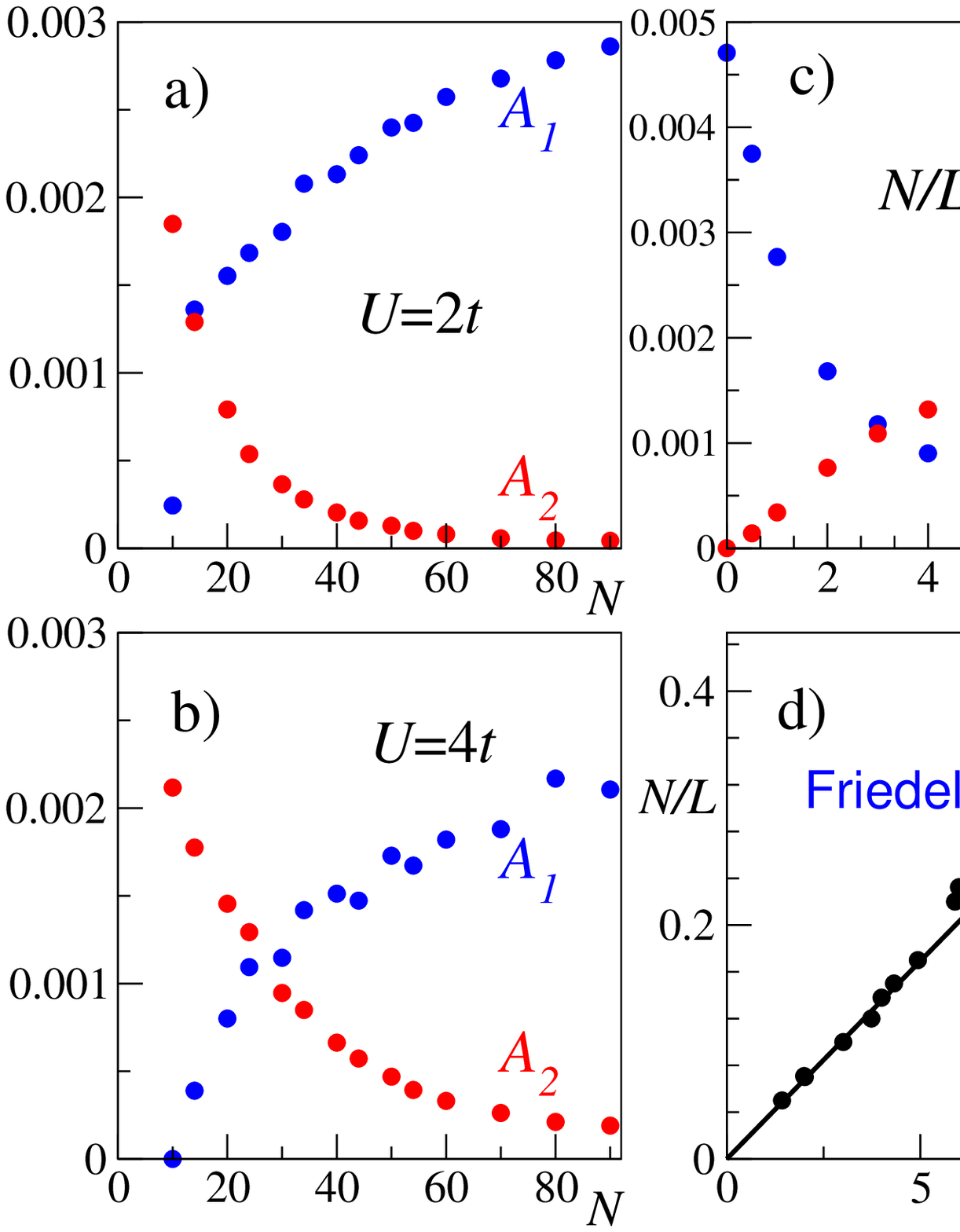}
   \end{center}
    \caption{Crossover of the amplitudes in Eq.~(\ref{rho_fit})
(error $\alt 10\% $) for $L=200$
from the DMRG data as a function of filling and interaction.
d) Crossover points ($A_1=A_2$) in the
$U$-$N/L$-plane showing the scaling with $N/LU \approx 0.034$ for $L=200$. The solid line 
separates the Wigner and Friedel regions.}
\label{amplitudes}
\end{figure} 

Figure \ref{density} shows the quality of typical fits to the DMRG data.
The oscillations in the middle of the chain are very well represented by the 
analytical expression (\ref{rho_fit}), 
while there are small deviations near the edges. Deviations
from Luttinger liquid theory near boundaries have also 
been observed before in the context of the local density of states.\cite{schneider}
In order to determine the asymptotic amplitudes $A_1$ and
$A_2$ in the middle of the chain as accurately as possible we have therefore excluded 
the first few sites near the ends in the fits.
The fits are sensitive enough to even
confirm the exact values of the wave-vectors $2\kf$ and $4\kf -\dk$, 
since small deviations of
order $\dk$ already would make the quality of the fits considerably worse.

The results for the Friedel amplitude $A_1$ and the Wigner amplitude $A_2$
are shown in Fig.~\ref{amplitudes} for $L=200$.  
The amplitudes
show a clear crossover from Friedel oscillations to Wigner crystal waves at 
low filling.  Interestingly, the Friedel oscillations are suppressed exactly when the 
Wigner crystal waves are strong and vice versa.  Therefore, it is possible to identify 
two distinct ''regions'' of Wigner and  Friedel behavior.  

{}From the Luttinger liquid theory it is not {\it a priori} obvious why 
Friedel and Wigner oscillations cannot be strong simultaneously, but 
from the discussion in Section \ref{dens} it is clear that the Friedel oscillations
must compete with the large Fermi sea in Fig.~\ref{filling} since both states
cannot be realized at the same time.

In Fig.~\ref{amplitudes}-d we have plotted the parameters for which the two 
amplitudes are equal $A_1=A_2$ for a given length of $L=200$, which we define
as the line at which the crossover between the two regions occurs.  Interestingly,
the crossover occurs along a line of constant
$N/UL$ for small fillings (e.g. $N/UL  \approx 0.034$ for $L=200$).  

The Wigner type behavior always occurs at low filling or equivalently at large $U$.
At first sight it appears rather counter-intuitive that the on-site interaction $U$
should show a stronger effect as the average inter-particle distance $L/N$ is increased.
This behavior is special to one dimension since the total kinetic energy scales with 
$(N/L)^3$ at low filling, which becomes always smaller
than the total interaction energy,  which
scales with $(N/L)^2$ as $N/L\to 0$. 

{}The Bethe Ansatz equations for the Hubbard model also show  scaling behavior in that limit. 
In particular, we find that the 
Luttinger parameter is given by the simple expression 
\be K_c \approx 0.5 + \left(\tfrac{N t}{U L}\right) 4 \ln\! 2 \, 
+{\cal O}(\tfrac{N^2}{L^2},\tfrac{t^2}{U^2})\label{kc} \ee  
in the limit of  low filling and large $U$.  This is shown in Fig.~\ref{ba}.

Numerically the scaling behavior for the crossover points is observed for 
each length $L$ separately.  However, if the slope of the crossover line
$N/UL$ is plotted as a function of length $L$ we observe a clear downturn in the limit of 
large $L$ as shown in Fig.~\ref{phase}. 

\section{Conclusions}\label{concl}
The observed crossover from Friedel to Wigner oscillations is not a true phase
transition, which cannot occur in the 1D Hubbard model.  
Nonetheless, the two regions show clearly different physical behavior.
In particular, in the region of $4\kf$ oscillations, the system is characterized
by {\it spin-incoherent} behavior with rather small or 
vanishing spin correlations according
to the arrangement in Fig.~\ref{filling}-b.  
As outlined in recent works the spin-incoherent Luttinger liquid 
shows significantly different
physical behavior, especially in regards to 
transport and tunneling characteristics.\cite{matveev}
The dominance of $4 \kf$ oscillations in finite systems 
appears to be an additional indicator for the onset 
of spin-incoherent behavior. 
The particular correlation functions for the local density in finite systems
are in fact well suited to study this crossover,
since the $4\kf$ term in Eq.~(\ref{4kF}) contains only charge
degrees of freedom, while the Friedel term in Eq.~(\ref{2kF})
also contains the spin boson $\phi_s$,
the amplitude of which is accordingly suppressed in  spin-incoherent regime.

\begin{figure}
   \begin{center}
        \includegraphics[width=0.50\textwidth,angle=0]{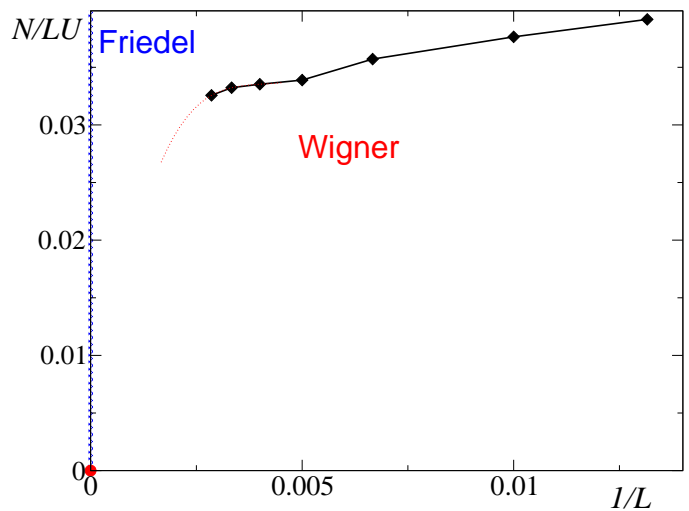}
   \end{center}
    \caption{Crossover 
 points $N/LU$ as a function of inverse length $1/L$.}
\label{phase}
\end{figure}

The length dependence of the crossover points $UL/N$
in Fig.~\ref{phase} violates scaling behavior:
For a given density $N/L$ and interaction strength $U$
all parameters in the Luttinger liquid theory 
($v_c$, $v_s$, $K_c$) are fixed. 
Nonetheless, it is still possible to observe a crossover from Wigner to Friedel 
oscillations as a function of length (moving horizontally in Fig.~\ref{phase}).
It is quite surprising that a critical model can cross over to different 
physical behavior as a function of length only, when all relevant parameters
are fixed.  
This remarkable violation of scale invariance is not due to higher 
order operators, such as the well-known logarithmic terms which also 
give strong corrections to the scaling behavior.\cite{schulz0}
Instead the crossover can be explained by  
the competition between vastly different velocities in 
the spin and the charge sector.
Naively, the onset of spin-incoherent Luttinger liquid 
would be expected 
 when the ultraviolet spin cutoff $v_s/a$ becomes comparable to
the infrared charge cutoff  $v_c/L$ giving a length dependent 
crossover.  While this may explain the 
broken scale invariance, this argument does not explain the
crossover line in Fig.~\ref{phase} quantitatively.
From the diagram alone it is also not clear that the Friedel region is always
stable in the thermodynamic limit $L\to \infty$, but a downturn as indicated
by the red dotted line would be expected for larger lengths $L$.

The arguments presented here are also valid for other 
Luttinger liquid systems, since the 
competition of spin and charge energy scales is generically always possible 
as a function of interaction and filling. 
Therefore, the crossover between the
different density oscillations discussed above
is a common signature of spin-charge separation in one dimension.

{}From the experimental side, Luttinger liquid behavior has so far only been
seen in very special cases, such as carbon nanotubes \cite{Bockrath1999,Lee2004}
or cleaved edge overgrowth wires.\cite{Auslaender2002}
There is some hope now that Luttinger liquid physics can also be realized 
with ultra-cold fermionic atoms in nearly ideal geometries formed by optical 
traps,\cite{demler} which would have the advantage that the density distribution
discussed here could in principle
be detected directly using high resolution cameras, electron beams,\cite{ott} or
noise interference.\cite{greiner}
Fermionic gases can already be cooled down to 
less than $1/10$ of the Fermi energy. The finite temperature will 
lead to a faster decay of the oscillations from the edges
that can be accounted for in the theory.\cite{modecalc} In fact it would be interesting to 
perform experiments in a regime where all spin excitations are smaller than the temperature. 
This would be a perfect realization of the spin incoherent Luttinger liquid, 
leading to a complete vanishing of the Friedel oscillations while the Wigner 
oscillations remain.
Hard edges can be implemented by focused laser beams or trapped 
impurity atoms.

In summary, we have systematically analyzed the local density distribution 
in finite Hubbard chains
as a function of 
filling, interaction strength $U$, and system size. 
A combination of bosonization and DMRG 
calculations allowed a detailed description of the density oscillations
in terms of the quantitative formula (\ref{rho_fit}).
For small interactions and large fillings $2\kf$ Friedel oscillations $A_1$
dominate, while the Wigner crystal amplitude $A_2$ remains small.
However, for smaller filling or increasing interactions  the
overall amplitude $A_1$ of the Friedel oscillations is strongly reduced
while $A_2$ grows.
This signals the crossover to a different physical region, which is 
described by a spin-incoherent large Fermi sea
with no double occupancy of spin up and down fermions.
The density oscillations 
we have described here are an accessible feature to study the 
crossover towards the spin-incoherent Luttinger liquid
in detail, e.g.~using ultra cold fermionic gases in 
1D optical traps. 

{\em Acknowledgments} We are thankful for useful discussions with Sebastian Reyes.
This work was supported by 
the DFG and the State of Rheinland-Pfalz via
the SFB/Transregio 49 and the MATCOR school of
excellence.



\begin{thebibliography}{99}
\bibitem{footnote}
The wave vector shift $-\dk x$ is taken from Eq.~(\ref{wigner2}).
\end{thebibliography}
\end{document}